\documentclass[twocolumn,floatfix]{revtex4}
\usepackage{graphicx}
\usepackage{dcolumn}
\usepackage{bm}
\usepackage{amsmath}
\usepackage{amsfonts}
\usepackage{amssymb}
\usepackage{color}
\usepackage{epstopdf}
\usepackage{float}
\providecommand{\U}[1]{\protect\rule{.1in}{.1in}}
\newcommand{\ket}[1]{|#1\rangle}
\newcommand{\bra}[1]{\langle#1|}

\begin{document}
\title{Temperature of a finite-dimensional quantum system}
\author{Andr\'es Vallejo}
\author{Alejandro Romanelli}
\author{Ra\'ul Donangelo}
\affiliation{\begin{small} Facultad de Ingenier\'{\i}a, Universidad 
de la Rep\'ublica, Montevideo, Uruguay\end{small}}
\date{\today}
\begin{abstract}
\noindent A general expression for the temperature of a 
finite-dimensional quantum system is deduced from thermodynamic arguments.
At equilibrium, this magnitude coincides with the standard thermodynamic
temperature. 
Furthermore, it is well-defined even far from equilibrium. 
Explicit formulas for the temperature of two and three-dimensional 
quantum systems are presented, and some additional relevant aspects 
of this quantity are discussed.

\end{abstract}

\maketitle

\section{\label{sec:level1}Introduction}

One of the main concerns of quantum thermodynamics is the correct
identification of quantum analogues of relevant thermodynamic magnitudes. 
Our experience with the macroscopic world induces us to express the
results of our research in this area in terms of concepts like
\textit{free energy}, \textit{work}, or \textit{entropy production}, 
which are well-defined in the classical world.
However, the connection between both worlds (classical and quantum) 
has been shown  not to be trivial, often counterintuitive, and in 
some cases, traumatic \cite{Einstein}.
As a consequence, after almost a century of existence of quantum 
theory, and despite the enormous progress of quantum thermodynamics 
in recent years \cite{Brandao,Parrondo,DAlessio,Goold,Vinjanampathy,Binder1}, 
most classical thermodynamic magnitudes are still waiting for an
universally accepted quantum definition. 

As an example, let us examine the two basic mechanisms of energy 
exchange: heat and work. 
Early works in quantum thermodynamics were based on the idea that 
heat is related to changes in the state $\rho$ of the system, 
while work is linked to the time variation of its Hamiltonian $H$,
controllable by the experimenter \cite{Alicki}. 
This point of view leads to the following straightforward partition 
of the energy change, $d\langle H\rangle$:
\begin{equation}\label{dE}
d\langle H\rangle=\text{tr}[H d\rho]+\text{tr}[\rho dH]
\end{equation}
where $\text{tr}$ denotes the trace operation. 
Under this perspective, the two terms on the r.h.s. of Eq. (\ref{dE}) 
can be clearly identified as heat and work, respectively, so the 
above equation can be considered to be an statement of the first 
law of thermodynamics.

This point of view has been widely employed during the last four 
decades, leading to very interesting results in matters such as 
the characterization of entropy production \cite{Deffner}, or 
the deduction of the Landauer limit \cite{Alicki2}. 
Nevertheless, several alternative definitions of heat and work have 
been used \cite{Bender,Jarzynski,Uzdin,Bera}. 
In particular, very recently two independent groups placed attention 
on the fact that certain changes in the state of the system are not 
accompanied by an entropy change, and argue that this fact 
implies that the part of the term $\text{tr}[H d\rho]$
which is not related to the change in the eigenvalues of $\rho$, but
in its eigenvectors, should not be considered as heat, but as work. 
As a consequence, new definitions of both magnitudes were proposed 
\cite{Alipour1,Ahmadi}.

Considering this new perspective, in this work we focus on the study 
of the concept of temperature, which, except for a few special 
cases, has been elusive to an extension to the quantum regime. 
Our approach is based of the fact that both entropy and energy 
have well accepted definitions in the quantum case, so the 
application of the standard definition of temperature,

\begin{equation}\label{tempSM}
\frac{1}{T}=\frac{\partial S}{\partial E}
\end{equation} 
\noindent is, at least in principle, plausible. 

This will also lead to the concept of temperature for a system 
out of equilibrium.
Although temperature is clearly an equilibrium property, ``effective temperatures"  and similar concepts have proven to be very useful 
in their respective contexts \cite{Gemmer,Johal,Poliblanc,Williams,Latune,Ali,Brunner2}, 
so the idea of an out-of-equilibrium temperature should 
not be discarded without a more profound exploration. 
One step in that direction is given in the present work. 

The remainder of this paper is organized as follows. 
In Section II, we apply Eq. (\ref{tempSM}) in order to find the 
temperature of a two-level system. 
This particular case is interesting in its own right, and 
gives some insight on the general situation.
In Section III, we deduce an expression for the temperature 
of a generic $N$-dimensional quantum system. 
Using this result, an explicit formula for the temperature of a qutrit 
is obtained, and shown to be consistent with the classical case.
Conclusions and final remarks are presented in Section IV.

\section{two-level system}
\subsection{Temperature}
In order to find the temperature of the system, we start by assuming 
that the von Neumann entropy $S_{vN}$ is a valid extension of the
thermodynamic entropy in the quantum regime.
This is based on the fact that in, thermal equilibrium, both entropies
coincide.
For a two-level system, $S_{vN}$ can be expressed in terms of the 
natural populations, i.e., the eigenvalues 
$\lbrace\lambda_{1},\lambda_{2}\rbrace$ of the density matrix of 
the system, as \cite{Breuer}:

\begin{equation}\label{SvN2level}
S_{vN}=-\lambda_{1}\text{ln}{\lambda_{1}}
       -\lambda_{2}\text{ln}{\lambda_{2}},
\end{equation}
Then, applying the usual definition of temperature (using $S_{vN}$ 
instead of the thermodynamic entropy $S$), we obtain:
\begin{equation}\label{T1}
\frac{1}{k_{B}T}=\frac{\partial S_{vN}}{\partial E}=\frac{\partial S_{vn}}
{\partial \lambda_{1}}\frac{\partial \lambda_{1}}{\partial E}+
\frac{\partial S_{vn}}{\partial \lambda_{2}}\frac{\partial \lambda_{2}}
{\partial E}.
\end{equation}
Evaluation of Eq. (\ref{T1}) requires to express the natural populations 
in terms of the energy, which, as usual, is defined as the expected 
value of the local Hamiltonian in the actual state:
\begin{equation}\label{energy1}
E=\langle H_{S}\rangle=\text{tr}[\rho_{S}H_{S}].
\end{equation} 
\noindent Expanding the density matrix $\rho_{S}$ in its instantaneous 
eigenbasis $\lbrace\ket{\psi_{1}},\ket{\psi_{2}}\rbrace$:
\begin{equation}\label{rhoS2level}
\rho_{S}=\lambda_{1}\ket{\psi_{1}}\bra{\psi_{1}}+
\lambda_{2}\ket{\psi_{2}}\bra{\psi_{2}},
\end{equation}
\noindent replacing Eq. (\ref{rhoS2level}) in Eq. (\ref{energy1}), 
and using that $\text{tr}[\ket{\psi_{j}}\bra{\psi_{j}}H_{S}]=
\bra{\psi_{j}}H\ket{\psi_{j}}$, we obtain:
\begin{equation}\label{energy2}
E=\lambda_{1}\bra{\psi_{1}}H_{S}\ket{\psi_{1}}+
\lambda_{2}\bra{\psi_{2}}H_{S}\ket{\psi_{2}}
\end{equation}

Eq. (\ref{energy2}) together with the fact that the density matrix
has a trace equal to one
\begin{equation}
\text{tr}(\rho_{S})=\lambda_{1}+\lambda_{2}=1,
\end{equation}
\noindent allows us to find $\lambda_{1}$ as a function of 
the energy:
\begin{equation}\label{langa12}
\lambda_{1}=\frac{E-\bra{\psi_{2}}H_{S}\ket{\psi_{2}}}
{\bra{\psi_{1}}H_{S}\ket{\psi_{1}}-\bra{\psi_{2}}H_{S}\ket{\psi_{2}}},
\end{equation}
\noindent and from the above equation we obtain
\begin{equation}\label{dLdE}
\frac{\partial \lambda_{1}}{\partial E}=\frac{1}{\bra{\psi_{1}}H_{S}
\ket{\psi_{1}}-\bra{\psi_{2}}H_{S}\ket{\psi_{2}}},
\end{equation}
\noindent and a similar expression for $\frac{\partial \lambda_{2}}{\partial E}$. 
On the other hand:
\begin{equation}\label{dSdL}
\frac{\partial S_{vN}}{\partial \lambda_{i}}=-(\text{ln}
{\lambda_{i}}+1),\; i=1,2.
\end{equation}

From Eqs. (\ref{T1}), (\ref{dLdE}) and (\ref{dSdL}), we obtain the final expression:
\begin{equation}\label{T2level}
T=\frac{\bra{\psi_{1}}H_{S}\ket{\psi_{1}}-\bra{\psi_{2}}H_{S}
\ket{\psi_{2}}}{k_{B}\text{ln}{\left(\lambda_{2}/\lambda_{1}\right)}}.
\end{equation}

\subsection{Discussion}
We first compare the above result with the well-known relation between 
the temperature and the populations in thermal equilibrium $P_{g}$ 
and $P_{e}$ of the ground and excited states, respectively, with
corresponding energies $E_{g}$ and $E_{e}$: 
\begin{equation}\label{T2eq}
T=\frac{E_{e}-E_{g}}{k_B \text{ln}(P_{g}/P_{e})}
\end{equation} 
\noindent Since in this particular case the density matrix and the 
Hamiltonian commute, both are diagonal in the same basis, and, as 
a consequence the above expressions become identical. 
Thus Eq. (\ref{T2level}) is a natural extension of Eq. (\ref{T2eq}) 
for the out-of-equilibrium case.
 
Let us now consider the (possibly time-dependent) Hamiltonian: 
\begin{equation}\label{Hs}
H_{S}=\varepsilon\ket{e}\bra{e}-\varepsilon\ket{g}\bra{g}.
\end{equation}
Using Eq. (\ref{Hs}), the terms in the numerator of Eq. (\ref{T2level}) are:
\begin{equation}\label{Cond1}
\bra{\psi_{i}}H_{S}\ket{\psi_{i}}=\varepsilon 
[\vert\bra{\psi_{i}}e\rangle\vert^{2}-
\vert\bra{\psi_{i}}g\rangle\vert^{2}], \; i=1,2.
\end{equation}
The density operator, Eq. (6), can also be expressed in the energy 
basis in terms of the components of the Bloch vector $\vec{B}=(u,v,w)$, 
which are a measure of the magnetization of the system, since they are 
proportional to the expected values of the spin operators. 
The expression is \cite{Nielsen}:
\begin{equation}\label{rhoS2level2}
\begin{split}
\rho_{S}=&\frac{1}{2}[(1+w)\ket{g}\bra{g}+(u-iv)\ket{g}\bra{e}\\
&+(u+iv)\ket{e}\bra{g}+(1-w)\ket{e}\bra{e}].
\end{split}
\end{equation}

It is easy to see that the eigenvalues of $\rho_{S}$ can be expressed
in terms of the modulus of the Bloch vector $B=(u^2+v^2+w^2)^{1/2}$ as:
\begin{equation}\label{eigenavlues}
\lambda_{1/2}=\frac{1}{2}(1\pm B).
\end{equation}
Using this result and Eqs. (\ref{rhoS2level}) and (\ref{rhoS2level2}), we obtain:
\begin{equation}\label{Cond2}
\begin{cases}
\vert\bra{\psi_{1}}g\rangle\vert^{2}-
\vert\bra{\psi_{2}}g\rangle\vert^{2}=\dfrac{w}{B}\\
\vert\bra{\psi_{1}}e\rangle\vert^{2}-
\vert\bra{\psi_{2}}e\rangle\vert^{2}=-\dfrac{w}{B}
\end{cases}
\end{equation}
Finally, from Eqs. (\ref{T2level}), (\ref{Cond1}) and (\ref{Cond2}), we obtain a
compact expression for the temperature of the two-level system:

\begin{equation}\label{T2level2}
T=\dfrac{\varepsilon w}{k_{B}B\tanh^{-1}(B)},
\end{equation}
\noindent which is discussed in more detail in Appendix A, together
with other properties of the system that emerge from this equation.

\section{General case}
\subsection{Temperature}
In this section, we will show that it is possible to generalize 
Eq. (\ref{T2level}) to any arbitrary finite dimension. 
First, let us consider what happens in the classical case.

In classical equilibrium thermodynamics, the macrostate of a system 
is determined when the values of a reduced set of extensive 
quantities are known. 
For some systems, a common selection of these quantities is the 
triplet $E,V,N$ (energy, volume, and particle number). 
This means that any other property is a function of this 
fundamental set. 

In the quantum case, the state of the system is determined by the 
density matrix, which, expressed in its natural basis is:
\begin{equation}
\rho_{S}= \sum_{j=1}^{N}\lambda_{j}\ket{\psi_{j}}\bra{\psi_{j}}.
\end{equation}
This means that given its eigenvectors, and assuming that their 
change is not related to heat transfer, a set of $N$ parameters 
is required to determine the state. 
Taking into account the normalization condition, and the restriction 
imposed by the energy
\begin{equation}\label{energy3}
E=\sum_{j=1}^{N}\lambda_{j}\bra{\psi_{j}}H_{S}\ket{\psi_{j}},
\end{equation} 
\noindent it is clear that $N-2$ independent quantities 
should be specified to define the quantum state. 

A second observation is that in the classical case, temperature 
is defined as the inverse of the partial derivative of the entropy 
with respect to the energy, keeping the other quantities of the 
fundamental set fixed. 
Since in the quantum case these quantities are not specified, and 
in order to keep the discussion as general as possible, we shall 
consider a set of $N-2$ observables 
\begin{small}$ \lbrace O^{1},...,O^{N-2}\rbrace$\end{small}, 
such that their expected values, together with the energy and the normalization condition, determine the state, and which will play 
the role of the complementary thermodynamic quantities that will 
be kept constant when defining the temperature. 
We should note that in the two-level case there was no need to 
introduce such additional quantities because the energy and the 
eigenvectors of $\rho_{S}$, suffice to completely determine the state. 
The expected values of these observables are,
\begin{equation}\label{Ok}
\langle O^{k}\rangle=\sum_{j=1}^{N}\lambda_{j}\bra{\psi_{j}}O^{k}\ket{\psi_{j}},
\hspace{0.2cm},\; k=1,..,N-2.
\end{equation} 
To simplify the notation, in what follows we denote the matrix 
elements of a local operator $X$ in the diagonal basis of $\rho_{S}$ 
as $\bra{\psi_{i}}X\ket{\psi_{j}}=X_{ij}$. 
From Eqs. (\ref{energy3}), (\ref{Ok}) and the normalization condition, 
we see that we can find the natural populations $\lambda_{j}$ by
solving the linear system $M\Lambda =\Gamma$, with
\begin{equation}\label{M}
M=\begin{pmatrix}\vspace{0.3cm}
{H_{11}}&{H_{22}}&{...}&{H_{NN}}\\\vspace{0.3cm}
{O^{1}_{11}}&{O^{1}_{22}}&{...}&{O^{1}_{NN}}\\\vspace{0.3cm}
{O^{2}_{11}}&{O^{2}_{22}}&{...}&{O^{2}_{NN}}\\\vspace{0.3cm}
{...}&{...}&{...}&{...}\\\vspace{0.3cm}
{O^{N-2}_{11}}&{O^{N-2}_{22}}&{...}&{O^{N-2}_{NN}}\\
{1}&{1}&{...}&{1}
\end{pmatrix},
\end{equation}
and $\Lambda =(\lambda_{1},...,\lambda_{N})^T$, 
$\Gamma=(E,\langle O^{1}\rangle,...,\langle O^{N-2}\rangle,1)^{T}$.
If $M$ satisfies $\text{det}(M)\neq0$, the system has the solution
\begin{equation}
\Lambda =M^{-1}\Gamma.
\end{equation}
Therefore the natural populations are lineal functions of the thermodynamic quantities, plus a constant term:
\begin{equation}
\lambda_{j}=M^{-1}_{j1}E+\sum_{i=2}^{N-1}M^{-1}_{ji}\langle O^{i-1}\rangle+M^{-1}_{jN}.
\end{equation}
It is now possible to obtain the temperature. 
The von Neumann entropy is given, in the present case, by
\begin{equation}\label{Svn2}
S_{vN}=-\sum_{j=1}^{N}\lambda_{j}\text{ln}{\lambda_{j}},
\end{equation} 
so we can define the temperature as
\begin{equation}\label{Tn}
\begin{split}
\frac{1}{k_{B}T}=&\frac{\partial S_{vN}}
{\partial E}\biggr\rvert_{\langle O^{1}\rangle ,...,
	\langle O^{N-2}\rangle}\\=&\sum_{j}
\frac{\partial S_{vN}}{\partial\lambda_{j}}\biggr\rvert_{\lambda_{k}
	\neq\lambda_{j}}\frac{\partial\lambda_{j}}
{\partial E}\biggr\rvert_{\langle O^{1}\rangle ,...,
	\langle O^{N-2}\rangle}
\end{split}
\end{equation}

The partial derivatives needed are
\begin{equation}\label{dSdL2}
\frac{\partial S_{vN}}{\partial\lambda_{j}}\biggr\rvert_{\lambda_{k}\neq\lambda_{j}}
=-(\text{ln}{\lambda_{j}}+1),
\end{equation}
and
\begin{equation}\label{dLdE2}
\frac{\partial\lambda_{j}}{\partial E}\biggr\rvert_{\langle O^{1}\rangle
	 ,...,\langle O^{N-2}\rangle}=M^{-1}_{j1}
\end{equation}
From Eqs. (\ref{Tn}), (\ref{dSdL2}) and (\ref{dLdE2}), we obtain
\begin{equation}\label{deftemp1}
T=-\left[k_{B}\sum_{j=1}^{N}M^{-1}_{j1}\left(\text{ln}\lambda_{j}
+1\right)\right]^{-1}
\end{equation}
A further simplification is possible. 
Since the last row of $M$ has equal elements, each one of the first 
$N-1$ columns of $M^{-1}$ verifies that the sum of its entries is zero.
Thus, $\sum_{j=1}^{N}M^{-1}_{j1}=0$, and 
\begin{equation}\label{Tgeneral}
T=-\frac{1}{k_{B}\sum_{j=1}^{N}M^{-1}_{j1}\text{ln}\lambda_{j}},
\end{equation}
which is the main result of this paper. 

As an example, we can write an explicit formula for the temperature 
of a three-level system.
Once the second relevant observable $O$ is selected, in order 
to find the temperature we must invert the matrix 
\begin{equation}
M=\begin{pmatrix}\vspace{0.3cm}
{H_{11}}&{H_{22}}&{H_{33}}\\\vspace{0.3cm}
{O_{11}}&{O_{22}}&{O_{33}}\\\vspace{0.3cm}
{1}&{1}&{1}
\end{pmatrix},
\end{equation}
It is straightforward to see that its inverse exists provided that
the condition
\begin{small}
\begin{equation}\label{determinant}
\text{det}(M)=H_{11}(O_{22}-O_{33})+H_{22}(O_{33}-O_{11})+
H_{33}(O_{11}-O_{22})\neq 0
\end{equation}
\end{small}
is satisfied.
If we now consider the first column of $M^{-1}$,
\begin{equation}\label{dL/dE3}
\begin{cases}\vspace{0.3cm}
M^{-1}_{11}=[O_{22}-O_{33}][\text{det}(M)]^{-1},\\\vspace{0.3cm}
M^{-1}_{21}=[O_{33}-O_{11}][\text{det}(M)]^{-1},\\
M^{-1}_{31}=[O_{11}-O_{22}][\text{det}(M)]^{-1},
\end{cases}
\end{equation}
\noindent from Eqs. (\ref{Tgeneral}), (\ref{determinant}) and (\ref{dL/dE3}), 
we obtain an explicit  expression for the temperature of a 
three-level system:

\begin{small}
\begin{equation}\label{T3}
T=
\frac{O_{11}(H_{33}-H_{22})+O_{22}(H_{11}-H_{33})+O_{33}(H_{22}-H_{11})}
{k_{B}[O_{11}\text{ln}(\lambda_{2}/\lambda_{3})+
 O_{22}\text{ln}(\lambda_{3}/\lambda_{1})+
 O_{33}\text{ln}(\lambda_{1}/\lambda_{2})]}
\end{equation}
\end{small}
\noindent where the eigenvalues $\lambda_{i}$ are
\begin{small}
\begin{equation}\label{eigenvalues3level}
\begin{cases}\vspace{0.3cm}
\lambda_{1}=\dfrac{H_{22}O_{33}-H_{33}O_{22}+E(O_{22}-O_{33})+\langle O\rangle (H_{33}-H_{22})}{\text{det}(M)},\\
\vspace{0.3cm} \lambda_{2}=\dfrac{H_{33}O_{11}-H_{11}O_{33}+E(O_{33}-O_{11})+\langle O\rangle(H_{11}-H_{33})}{\text{det}(M)}\\
\lambda_{3}=\dfrac{H_{11}O_{22}-H_{22}O_{11}+E(O_{11}-O_{22})+\langle O\rangle (H_{22}-H_{11})}{\text{det}(M)},
\end{cases}
\end{equation}
\end{small}
\subsection{Application}
As a concrete application, let us consider the particular case of a
discrete-time quantum walk on the line with three internal states 
\cite{Inui}.

Quantum walks are the quantum counterpart of classical random walks, 
and they are fundamental tools in areas such as quantum computation 
and quantum simulation 
\cite{Ambainis,Schreiber,Venegas,Portugal,Di Molfetta,Arrighi}.
The system evolves in the composite Hilbert space
$\mathcal{H}_{n}\otimes\mathcal{H}_{_{S}}$,
where $\mathcal{H}_{n}$ is the position space spanned by the basis
$\lbrace\vert n\rangle\rbrace$, and  $\mathcal{H}_{_{S}}$ is the 
chirality space, which in this case has dimension three, and is 
spanned by the kets $\ket{R}$ (associated with steps to the right), $\ket{N}$ (stay in place), and $\ket{L}$ (steps to the left).
We consider the particular dynamics given by successive applications 
of the operator
\begin{equation}
U= \mathcal{T}(I^{n}\otimes G),
\end{equation}
where $G$ is the \textit{Grover coin}:
\begin{small}
\begin{equation}
G=\dfrac{1}{3}\begin{pmatrix}
{-1}&{2}&{2}\\
{2}&{-1}&{2}\\
{2}&{2}&{-1}
\end{pmatrix},
\end{equation}
\end{small}
$\mathcal{T}$ is the conditional shift operator:
\begin{equation}
\begin{split}
\mathcal{T}=&\sum_{n}\ket{n+1}\bra{n}\otimes\ket{R}\bra{R}+
\ket{n-1}\bra{n}\otimes\ket{L}\bra{L}\\
 &+\ket{n}\bra{n}\otimes\ket{N}\bra{N},
 \end{split}
\end{equation}
and $I^{n}$ is the identity operator in $\mathcal{H}_{n}$. 
Thus, the global state at time $t$ is given by
\begin{equation}
\ket{\psi(t)}=U^{t}\ket{\psi (0)},
\end{equation}
where $\ket{\psi(0)}$ can be written as
\begin{small}
\begin{equation}
\ket{\psi(0)}=\sum_{n}\ket{n} \otimes [a_{n}(0)\ket{R}
+b_{n}(0)\ket{N}+c_{n}(0)\ket{L}],
\end{equation}
\end{small}
\noindent with $a_{n}(0)$, $b_{n}(0)$ and $c_{n}(0)$ satisfying 
the normalization condition
$\sum_{n}\left(\vert a_{n}(0)\vert^{2}
+\vert b_{n}(0)\vert^{2}
+\vert c_{n}(0)\vert^{2}\right)=1$.

We now focus on the evolution of the coin, which can be considered 
as a three-level system in interaction with a large environment, 
represented by the position Hilbert space $\mathcal{H}_{n}$. 
To make sure that the environment is effectively large at any time, 
we restrict the study to Gaussian position distributions that span
a large number of sites, i.e. with a standard deviation $\sigma \gg 1$. 
For convenience we take equal initial amplitudes for the left and 
right states,
\begin{equation}\label{inistateQW}
\begin{cases}
a_{n}(0)=c_{n}(0)=\frac{e^{\frac{-n^{2}}{4\sigma^{2}}}}
{\sqrt[4]{2\pi\sigma^{2}}} a_{0}\\
b_{n}(0)= \frac{e^{\frac{-n^{2}}{4\sigma^{2}}}}
{\sqrt[4]{2\pi\sigma^{2}}}b_{0} .
\end{cases}
\end{equation}

As in the quantum walk with two internal states described 
in Refs. \cite{Vallejo0,Vallejo2}, in this case the open evolution of 
the coin, starting from the initial condition given by 
Eq. (\ref{inistateQW}), consists of an unitary part described 
by the effective local Hamiltonian $G$, and a dissipative 
contribution due to interactions with the environment. 
In particular, in the asymptotic regime, the coin reaches an 
equilibrium state characterized by its reduced density matrix $\rho^{\infty}_{S}$, which adopts the general form: 
\begin{equation}\label{rhoSQW}
\rho^{\infty}_{S}=\begin{pmatrix}\vspace{0.3cm}
{1/3}&{x}&{x}\\\vspace{0.3cm}
{x}&{1/3}&{x}\\\vspace{0.3cm}
{x}&{x}&{1/3} 
\end{pmatrix},
\end{equation}
\noindent where $x$ depends on the initial state, and, for 
physical states, it satisfies the inequality $-1/6<x<1/3$. 
On the other hand, a simple calculation shows that the 
thermal state associated to the local Hamiltonian $H_{S}=G$ 
at a given inverse temperature $\beta$, i.e. 
\begin{small}$e^{-\beta G}/\text{tr}[e^{-\beta G}]$\end{small}, 
has exactly the form of Eq. (\ref{rhoSQW}), with
\begin{equation}\label{x(beta)}
x=-\dfrac{2\sinh \beta}{3(3\cos\beta+\sinh\beta)}
\end{equation}
This implies that the equilibrium state can be considered a 
thermal state, and from the above equation we can obtain the 
corresponding equilibrium temperature ($k_{B}=1$):
\begin{equation}\label{TeqQW}
T_{eq}=\frac{1}{2}\text{ln}\left(\dfrac{1-3x}{1+6x}\right).
\end{equation} 

As the observable $O$, we choose one represented by one of the 
Gell-Man matrices, which, together with the identity, form 
a basis for the Hermitian operators in dimension three:
\begin{small}
$$
O_{1}=\begin{pmatrix}\vspace{0.3cm}
{0}&{1}&{0}\\\vspace{0.2cm}
{1}&{0}&{0}\\\vspace{0.2cm}
{0}&{0}&{0}
\end{pmatrix},\hspace{0.2cm}
O_{2}=\begin{pmatrix}\vspace{0.2cm}
{0}&{-i}&{0}\\\vspace{0.2cm}
{i}&{0}&{0}\\\vspace{0.2cm}
{0}&{0}&{0}
\end{pmatrix},
\hspace{0.2cm}
O_{3}=\begin{pmatrix}\vspace{0.2cm}
{1}&{0}&{0}\\\vspace{0.2cm}
{0}&{-1}&{0}\\\vspace{0.2cm}
{0}&{0}&{0}
\end{pmatrix}
$$
$$
O_{4}=\begin{pmatrix}\vspace{0.3cm}
{0}&{0}&{1}\\\vspace{0.2cm}
{0}&{0}&{0}\\\vspace{0.2cm}
{1}&{0}&{0}
\end{pmatrix},\hspace{0.2cm}
O_{5}=\begin{pmatrix}\vspace{0.2cm}
{0}&{0}&{-i}\\\vspace{0.2cm}
{0}&{0}&{0}\\\vspace{0.2cm}
{i}&{0}&{0}
\end{pmatrix},\\\hspace{0.2cm}
O_{6}=\begin{pmatrix}\vspace{0.2cm}
{0}&{0}&{0}\\\vspace{0.2cm}
{0}&{0}&{1}\\\vspace{0.2cm}
{0}&{1}&{0}
\end{pmatrix}
$$
$$
O_{7}=\begin{pmatrix}\vspace{0.3cm}
{0}&{0}&{0}\\\vspace{0.2cm}
{0}&{0}&{-i}\\\vspace{0.2cm}
{0}&{i}&{0}
\end{pmatrix},\hspace{0.2cm}
O_{8}=\dfrac{1}{\sqrt{3}}\begin{pmatrix}\vspace{0.2cm}
{1}&{0}&{0}\\\vspace{0.2cm}
{0}&{1}&{0}\\\vspace{0.2cm}
{0}&{0}&{-2}
\end{pmatrix},
$$
\end{small}

\begin{figure}
	\label{fig1}
\centering
{\includegraphics[width=1.0\columnwidth]{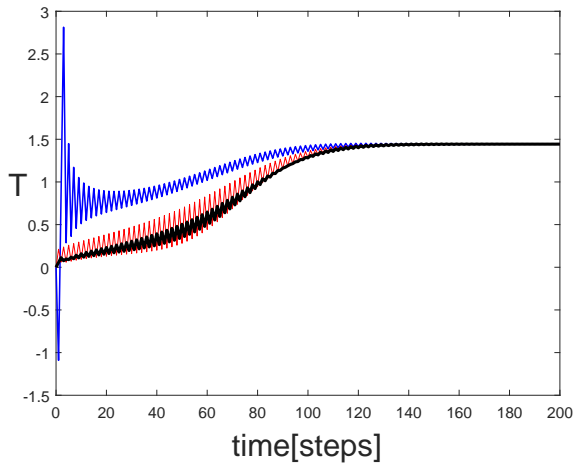}}
\caption{Evolution of the temperature associated to the 
observables $O_{1}$ and $O_{6}$ (blue, normal line), 
$O_{3}$ and $O_{8}$ (red, thin line), and $O_{4}$ 
(black, wide line). }
\end{figure}

The numerical analysis shows that  $O_{2}$, $O_{5}$ and $O_{7}$ 
are trivial constants of motion. 
Their expected value is zero during the entire evolution, 
and the corresponding diagonal matrix elements, and the 
determinant of the matrix $M$) are also zero, so entropy 
cannot be expressed in terms of the energy and any of 
these observables. 
For the rest of the set, the corresponding temperatures 
are shown in Fig. \ref{fig1}.
The observables $O_{1}$ and $O_{6}$ have exactly the same 
expected value during the entire evolution, leading to the 
same behavior of the temperature. 
The same occurs with the observables $O_{3}$ and $O_{8}$, 
as it can be seen in Fig. \ref{fig1}. 
We note that as time increases, thermal equilibrium is 
attained, and the temperatures become independent of the 
observable chosen, as they converge to a common asymptotic value. 
For the initial state selected we have that 
$x\simeq -0.1112$, so the asymptotic temperature can be found theoretically from Eq. (\ref{TeqQW}), obtaining,
\begin{equation}
T_{eq}\simeq 1.4418,
\end{equation}
which coincides with the limit value of Fig. \ref{fig1}. 
A general proof of the fact that in thermal equilibrium 
the temperature is independent of the observables $O$ 
employed in its definition is given in the following subsection.
  
\subsection{Discussion}
Eq. (\ref{Tgeneral}) depends on the eigenvalues of the 
density matrix, but also, through the parameters $M^{-1}_{j1}$, 
on the eigenstates. Thus, all the information contained 
in the state is required in order to find the temperature. 

The dependence on the generic observables 
\begin{small} $\lbrace O^{1},...,O^{N-2}\rbrace$\end{small} 
is an interesting characteristic, since it allows us to define 
\textit{a family of temperatures} that describes how the entropy 
varies with energy in different situations. 
Of course, the most appropriate selection of these observables,
i.e. those closest to their classical counterparts, should be
considered for the specific system under study.

A fundamental point about Eq. (\ref{Tgeneral}) is that, despite 
the need for introducing such properties to define the temperature, 
in thermal equilibrium the result is independent of the properties selected, a fact that can be shown as follows.

If thermal equilibrium is reached with an environment at temperature 
$T_{E}=1/\beta_{E}$, we have that:
\begin{equation}\label{thermalpop}
\lambda_{j}=\dfrac{e^{-\beta_{E} E_{j}}}{Z}
\end{equation}
where $Z$ is the partition function and $\{ E_{j}\}$ is the set of 
eigenenergies of the system Hamiltonian. 
Using Eq. (\ref{thermalpop}), and the fact that \begin{small}$\sum_{j=1}^{N}M^{-1}_{j1}=0$\end{small}, 
the denominator of Eq. (\ref{Tgeneral}) reads:
\begin{equation}\label{condition1}
\sum_{j=1}^{N}M^{-1}_{j1}\text{ln}(\lambda_{j})
 =-\beta_{E}\sum_{j=1}^{N}M^{-1}_{j1}E_{j}.
\end{equation}
On the other hand, since in thermal equilibrium 
$[H_{S},\rho_{S}]=0$, the instantaneous eigenstates of 
$\rho_{S}$ coincide with the energy eigenstates 
$\lbrace\ket{E_{i}}\rbrace$, so the diagonal elements 
of the Hamiltonian in the eigenbasis of $\rho_{S}$ are, 
in this case, the eigenenergies:
\begin{equation}
\bra{\psi_{j}}H_{S}\ket{\psi_{j}}=E_{j},
\end{equation}
so the first row of the matrix $M$ verifies that  
$M_{1j}=E_{j}$. 
As a consequence,
\begin{equation}\label{condition2}
\sum_{j=1}^{N}M^{-1}_{j1}E_{j}=1,
\end{equation}
Therefore, from 
Eqs. (\ref{Tgeneral}), (\ref{condition1}) and (\ref{condition2}),
we obtain,
\begin{equation}
T=T_{E},
\end{equation}
showing that the consistency of our temperature with the thermodynamic temperature in thermal equilibrium, is 
independent of the choice of the complementary properties.

Regarding the behavior of the temperature, note that as 
the state of system becomes closer to a pure state, $N-1$ 
of its eigenvalues tend to zero, so the denominator of 
Eq.(\ref{Tgeneral}) diverges and the temperature takes 
the value $T=0$. 
In the opposite case of a maximally mixed state, since the 
eigenvalues are equal, we can factorize $\text{ln}(1/N)$ 
in the denominator and the temperature diverges due to the 
condition \begin{small}$\sum_{j=1}^{N}M^{-1}_{j1}=0$\end{small}.

Finally, we would like to compare the present results with a 
related concept, the \textit{spectral temperature} \cite{Gemmer}, 
which is a commonly employed definition of temperature 
for out-of-equilibrium quantum systems \cite{Hartmann1,Hartmann2,Cai,Lin,Bhattacharya,Seveso,Walschaers}.

For a non-degenerate N-level system, the spectral temperature 
$\tau$ is defined as,
\begin{small}
\begin{equation}\label{spectral}
\frac{1}{\tau}=\left(1-\frac{P_{1}+
P_{N}}{2}\right)^{-1}\sum_{j=1}^{N-1}
\left[\frac{P_{j+1}+P_{j}}{2}\right]
\frac{\text{ln}(P_{j}/P_{j+1})}{E_{j+1}-E_{j}},
\end{equation} 
\end{small}
\noindent where $E_{j}$ are the eigenenergies, and $P_{j}$ the
corresponding probabilities under projective measurements in 
the energy eigenbasis. 
It can be shown that this expression, as it is the case with
the one we propose in this work, reduces to the ordinary 
temperature in thermal equilibrium. 
In addition, it has the advantage of being easy to calculate, 
unlike Eq. (\ref{Tgeneral}), which requires the inversion of a 
potentially large matrix. 
Nevertheless, a critical observation can be made considering 
the simple case of a two level system, situation in which 
Eq. (\ref{spectral}) becomes
\begin{equation}\label{spectral2}
\tau=\frac{E_{2}-E_{1}}{\ln(P_{1}/P_{2})}
\end{equation} 
Recalling Eq. (\ref{T2eq}), we see that this expression 
has the same structure than the equilibrium temperature, 
but considering the out-of-equilibrium populations. 
Although this strategy is commonly employed to assign a 
temperature to two-level systems \cite{Liao,Quan}, it is 
possible to see that it represents the change in entropy 
associated to a change in the energy only in some special cases. 
For instance, an unitary (isentropic) process in which 
the populations change in time can be implemented using 
a time-dependent Hamiltonian.  
In that case, the derivative of entropy with respect to 
energy is zero, while the spectral temperature is not. 
Our proposal does not suffer from this problem since, 
by construction, the temperature defined in this work  
is, precisely, the derivative of the entropy with respect 
to energy. 

This can be confirmed by analyzing the same situation but 
employing Eq. (\ref{T2level}), instead of Eq. (\ref{spectral2}). 
Since the system remains in a pure state, one of its eigenvalues 
is 1 and all others are 0,  and therefore the denominator 
diverges, leading to $T=0$. 
Furthermore, while the temperature in thermal equilibrium 
depends only on the populations, it would be reasonable 
to think that in the out-of-equilibrium situation, other 
aspects of the quantum state play a role.
This is so because the energy, Eq. (\ref{energy3}), is a 
function of the eigenvalues  and the eigenstates of $\rho_{S}$. 
Spectral temperature, as defined through Eq. (\ref{spectral}), 
does not meet this expected behavior.  

\section{Final Remarks and conclusions}
In this work we have shown that it is possible to define 
the notion of temperature for finite-dimensional quantum 
systems, in a way such that the consistency with the classical definition is preserved. 
Our proposal is based on the fact that the von Neumann entropy 
and the internal energy are well-defined functions of the state 
of the system, even far from equilibrium. 
We have presented explicit formulas for the temperature of
low-dimensional systems ($\text{dim}(\mathcal{H}_{S})=2,3$), but
similar expressions exist for higher dimensional systems.

An important aspect of our method is that, in order to define 
the temperature, one is required to select a set of observables 
such that their expected values are kept fixed when performing 
the partial derivative calculations. 
They play an analogous role to volume and particle numbers in 
classical thermodynamics, and this degree of freedom in the 
definition allows to adapt the application of the formula 
to the context, selecting those properties that are most 
convenient to keep fixed. 
Possible restrictions on the choice of these observables that 
ensure that the matrix M is invertible require a more profound 
study.
Further aspects of the present definition, such as its relation 
with the direction of the heat flow, are currently under investigation.

\section*{Acknowledgments}
This work was partially supported by Comisi\'on Acad\'emica de Posgrado (CAP), Agencia Nacional de Investigaci\'on e Innovaci\'on (ANII), and Programa de Desarrollo de las Ciencias B\'asicas (PEDECIBA) (Uruguay).

\appendix 
\section{Comments on Eq.(\ref{T2level2})}
In Ref. \cite{Vallejo1} an expression for the temperature of a 
two-level system similar to Eq. (\ref{T2level2}) was found, but 
in which the factor $w/B$ appears inverted. 
In order to help understand the origin of this difference, we first 
briefly present the arguments used in that work. 

We note that since the state of a two-level system is completely 
defined by the components  of the Bloch vector $(u,v,w)$, any 
thermodynamic property can be expressed in terms of these three components. 
In particular, as shown in Ref. \cite{Vallejo1}, the entropy is 
a function of the modulus $B$ of the Bloch vector:
\begin{equation}\label{entropyAp}
S_{vN}=
-\left(\frac{1+B}{2}\right)\ln\left(\frac{1+B}{2}\right)
-\left(\frac{1-B}{2}\right)\ln\left(\frac{1-B}{2}\right) .
\end{equation}

Choosing the $z$ axis in the direction of 
$\lbrace\ket{g},\ket{e}\rbrace$, the internal energy associated 
to the Hamiltonian (\ref{Hs}) is:
\begin{equation}\label{energyAp}
E=-\varepsilon B\cos\theta =-\varepsilon w
\end{equation}
where $\theta$ is the polar angle of $\vec{B}$ in spherical coordinates. 
Then, Ref. \cite{Vallejo1} defines the inverse temperature as the 
partial derivative of the entropy with respect to the energy in a 
zero work process from the standard point of view \cite{Alicki}, 
i.e. fixing the Hamiltonian, which in this case means fixing the 
$z$ axis and $\varepsilon$:
\begin{equation}
\frac{1}{k_{B}T}=\frac{\partial S_{vN}}{\partial E}\biggr{\vert}_{\varepsilon}
\end{equation}

Since entropy depends on the three components of the Bloch vector, 
but energy depends only on the $z$ component, applying the chain 
rule we obtain:
  
\begin{equation}\label{temp1Ap}
\frac{1}{k_{B}T}=\frac{\partial S_{vN}}{\partial B}
\frac{\partial B}{\partial w}
\frac{\partial w}{\partial E}\biggr{\vert}_{\varepsilon}
\end{equation} 

Finally, using Eqs. (\ref{entropyAp}) and (\ref{energyAp}), 
the expression for the temperature of the two-level obtained 
in reference \cite{Vallejo1} is:
\begin{equation}\label{temperature1}
T=\frac{\varepsilon B}{k_{B}w\tanh^{-1}(B)}
\end{equation}
Observe the different position of the factors $B$ and $w$ with 
respect to Eq. (\ref{T2level2}).

The previous argument was based on the hypothesis that the work 
done on the system is exclusively related to the changes in the 
local Hamiltonian. 
In the present work, we have adopted the new point of view 
presented in \cite{Alipour1,Ahmadi}. 
This new perspective can be illustrated by considering the 
infinitesimal energy change obtained by differentiating 
Eq. (\ref{energy2}):
\begin{equation}\label{dEAp}
\begin{split}
dE=&d\lambda_{1}\bra{\psi_{1}}H_{S}\ket{\psi_{1}}+
\lambda_{1}d\bra{\psi_{1}}H_{S}\ket{\psi_{1}}\\
&+d\lambda_{2}\bra{\psi_{2}}H_{S}\ket{\psi_{2}}+
\lambda_{2}d\bra{\psi_{2}}H_{S}\ket{\psi_{2}}
\end{split}
\end{equation}
Since entropy depends only on the eigenvalues of $\rho_{S}$, 
only the first and third terms of Eq. (\ref{dEAp}) contribute 
to the entropy change of the system. 
As a consequence, they are the only terms which should be 
considered as heat. 
This implies that work can be performed on the system even 
in the case that the Hamiltonian is kept fixed, as long as 
the eigenvectors of $\rho_{S}$ change in time.  

The adoption of this point of view implies that in the attempt 
of defining temperature by analogy with the classical case, 
the restriction of keeping the equivalent to the volume fixed 
in the partial derivative, i.e. zero work, implies that in 
the differentiation not only the Hamiltonian is kept fixed, 
as in Ref. \cite{Vallejo1}, but also the eigenstates of 
the reduced density matrix are kept fixed. 
Since an instantaneous eigenstate of $\rho_{S}$ in the Bloch 
sphere is defined by the pair of angles ($\theta , \varphi$), 
the definition of temperature from this new perspective becomes:

\begin{equation}
\dfrac{1}{k_{B}T}=\dfrac{\partial S_{vN}}{\partial E} 
\biggr{\vert}_{\varepsilon, \theta,\varphi}= 
\dfrac{dS_{vN}}{dB}\dfrac{\partial B}{\partial E} 
\biggr{\vert}_{\varepsilon, \theta,\varphi}
\end{equation}
The two factors in the above equation can be obtained from 
Eqs. (\ref{entropyAp}) and (\ref{energyAp}):
\begin{equation}
\dfrac{dS_{vN}}{dB}=-\tanh^{-1}(B),\hspace{0.3cm}
\dfrac{\partial B}{\partial E}\biggr{\vert}_{\varepsilon,\theta,\varphi} 
=-\dfrac{1}{\varepsilon\cos(\theta)}
\end{equation}
Finally, from the above equations and using that $\cos\theta =w/B$, 
we arrive at Eq. (\ref{T2level2}). 
It is clear that the adoption of Eq. (\ref{T2level2}) or 
Eq. (\ref{temperature1}) as the qubit's temperature is directly 
linked to a more fundamental question, which is, what are 
the most appropriate definitions of heat and work in the 
quantum regime? 
Perhaps the exploration of the corresponding temperatures can 
help to shed light on this matter.

In that sense, one fundamental difference between both approaches 
is related to the existence of an intrinsic entropy production. 
In \cite{Vallejo1} it was shown that the adoption of 
Eq. (\ref{temperature1}) as the qubit's temperature implies the 
existence of an entropy production which represents a measure 
of the loss of internal coherence by the qubit. 
In this new approach employed in this work, it is straightforward 
to verify that $dS_{vN}=\delta Q/T$, which implies no internal 
entropy production, as reported in \cite{Ahmadi2}. 
Of course, since in the general case the system is in an 
out-of-equilibrium state, the temperatures of the system and the 
environment may not coincide, so a boundary contribution of entropy 
production due to the possibly finite temperature difference is expected.
  
\begin{figure}[!h]
{\includegraphics[trim= 300 50 -50 0,scale=0.55,clip] {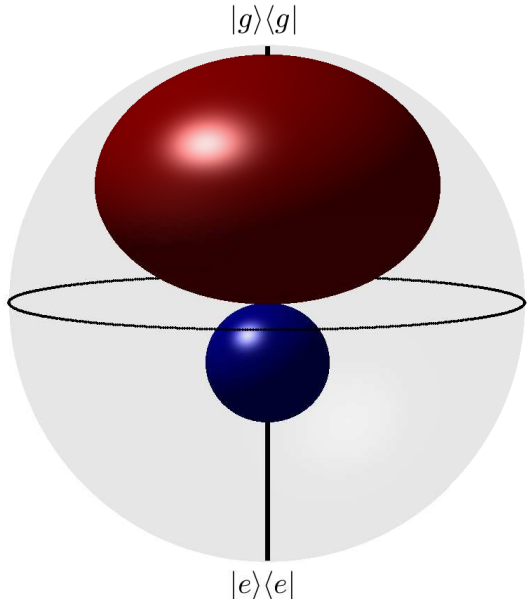}}
\caption{Isothermal surfaces in the Bloch sphere, corresponding to the
temperature values $k_{B}T_{1}=0.5\varepsilon$ (red, northern hemisphere) 
and $k_{B}T_{2}=-2\varepsilon$ (blue, southern hemisphere). }
\label{fig2}
\end{figure}

Equation (\ref{T2level2}) can be used to analyze the properties 
associated to the temperature obtained in the present approach. 
Note that, since $w$ is the component of the Bloch vector parallel 
to the effective magnetic field, the temperature defined in 
Eq. (\ref{T2level2}) is positive when the scalar product between 
the magnetization and the field is positive. 
On the other hand, we can see that pure states ($B=1$) and those 
with $w=0$ have zero temperature, except for the maximally mixed 
state ($B=0$), for which $T=\infty$. 
To illustrate this, two isothermal surfaces in the Bloch sphere,
one associated with a positive and another with a negative temperature,
are represented in Fig. (\ref{fig2}).

As an application of the concept of temperature of a two level system
employed in the present work, we will obtain the expression of 
the heat capacity of a qubit. 
In classical thermodynamics, the heat capacity is defined as 
the derivative of the energy with respect to temperature in a 
zero work process. 
Thus we define:
\begin{equation}\label{C}
C=\dfrac{\partial E}{\partial T}\biggr{\vert}
_{\varepsilon, \theta, \varphi}
\end{equation}
Using Eqs. (\ref{T2level2}) and (\ref{energyAp}), we have that,
\begin{equation}
E=-\varepsilon\cos\theta\tanh
\left(\dfrac{\varepsilon\cos\theta}{k_{B}T}\right)
\end{equation}
And therefore:
\begin{equation}\label{C2}
C=\left[\dfrac{\varepsilon\cos\theta/k_{B}T}
{\cosh\left(\varepsilon\cos\theta/k_{B}T\right)}\right]^{2}
\end{equation}
Clearly the heat capacity $C$ is non-negative, and since in 
thermal equilibrium the Bloch vector is parallel to the magnetic 
field, which is along the $z$-axis, and the temperature is equal 
to the equilibrium temperature $T_{E}$, Eq. (\ref{C2}) reduces to 
the classical expression for the heat capacity:
\begin{equation}
C=\left[\dfrac{\varepsilon/k_{B}T}
{\cosh\left(\varepsilon/k_{B}T\right)}\right]^{2}.
\end{equation}

\end{document}